\RequirePackage{fix-cm}
\documentclass[smallextended]{svjour3}       % onecolumn (second format)
\smartqed  % flush right qed marks, e.g. at end of proof
\usepackage{graphicx}
\begin{document}

\title{Resolution-scale relativistic formulation of non-differentiable mechanics%\thanks{Grants or other notes
%about the article that should go on the front page should be
%placed here. General acknowledgments should be placed at the end of the article.}
}
\subtitle{}

%\titlerunning{Short form of title}        % if too long for running head

\author{Mei-Hui Teh \and Laurent Nottale         \and Stephan LeBohec}

%\authorrunning{Short form of author list} % if too long for running head

\institute{M.-H. Teh \and S. LeBohec \at
Dept. of Physics and Astronomy, \\
201 James Fletcher Bldg. 115 S. 1400 E.\\
University of Utah\\
Salt Lake City, UT 84112 , USA   \\
\email{meihui.teh@gmail.com}\\           
\email{lebohec@physics.utah.edu}           
\and
L. Nottale \at
LUTH, Observatoire de Paris,\\ 
5, place Jules Janssen\\
 Meudon, 92195, France\\
\email{Laurent.nottale@obspm.fr}           
 }

\date{Received: date / Accepted: date}
% The correct dates will be entered by the editor

\maketitle

\begin{abstract}
This article motivates and presents the scale relativistic approach to non-differentiability in mechanics and its relation to quantum mechanics. It stems from the scale relativity proposal to extend the principle of relativity to resolution-scale transformations, which leads to considering non-differentiable dynamical paths.  We first define a complex {\it scale-covariant time-differential operator} and show that mechanics of non-differentiable paths is implemented in the same way as classical mechanics but with the replacement of the time derivative and velocity with the time-differential operator and associated complex velocity. With this, the generalized form of Newton's fundamental relation of dynamics is shown to take the form of a Langevin equation in the case of stationary motion characterized by a null average classical velocity. The numerical integration of the Langevin equation in the case of a harmonic oscillator taken as an example reveals the same statistics as the stationary solutions of the Schr\"odinger equation for the same problem. This motivates the rest of the paper, which shows Schr\"odinger's equation to be a reformulation of Newton's fundamental relation of dynamics as generalized to non-differentiable geometries and leads to an alternative interpretation of the other axioms of standard quantum mechanics in a coherent picture. This exercise validates the scale relativistic approach and, at the same time, it allows to envision macroscopic chaotic systems observed at resolution time-scales exceeding their horizon of predictability as candidates in which to search for {\it quantum-like}  dynamics and structures. 
% \PACS{PACS code1 \and PACS code2 \and more}
% \subclass{MSC code1 \and MSC code2 \and more}
\end{abstract}

\section{Introduction}\label{intro}
The relativity principle prescribes the laws of physics to have the same expression in all reference frames. It is implicitly understood that the relation between two reference frames is entirely specified by their relative position, orientation, and motion. The galilean implementation of the relativity principle with the least action principle  establish the whole of classical mechanics \cite{landauvol1}. The special relativistic implementation with the identification of the speed of light as an invariant under the Lorentz transformations of the coordinates allows for a simple and natural theory of electrodynamics. Galilean and special relativity apply to the class of inertial reference systems, all in uniform relative motion with respect to each other. The transformation of the coordinates from one reference system to another is then linear. The next step  consists in extending the relativity principle to non-inertial reference systems. Coordinates transformations are then not necessarily linear anymore but they are twice differentiable diffeomorphisms. The least action principle takes the form of the geodesic principle and, with the equivalence principle, this leads to considering gravitation as a manifestation of the curvature of space-time\cite{Weinberg1972,Fock1964}.

From a purely geometrical point of view, the next step toward greater generality would be to consider transformations that are still continuous but not necessarily differentiable. The task of implementing such a generalization of the relativity principle is absolutely formidable.  However, drawing from general relativity, one can envision  two direct  implications stemming from the abandonment of the differentiability hypothesis.   First, geodesics as dynamical paths can be expected to lose their enumerable nature. On a non-differentiable space, two points would be connected by an infinite number of paths. This is a situation in which determinism, to be implemented by a generalized principle of least action, has to give way to a probabilistic and statistical description. Second, paths remain continuous but are non-differentiable, implying their  resolution-scale dependent and divergent nature, that is fractal in a general meaning\cite{Mandelbrot1977}. This identifies scaling laws regarded as changes of resolution-scales as the essential notion for the implementation of a non-differentiable extension to relativity. This corresponds to including resolution-scales as additional relative specifications of reference frames and extending  the principle of relativity to resolution-scale transformations. This proposal, known as scale relativity, was originally formulated by one of the authors \cite{nottale1993,nottale2011}.   The present article constitutes a review of the scale relativitic approach for the development of a mechanics of non-differentiable dynamical paths  with the necessary explicit introduction of observations resolution-scales.  

The scale relativity approach can be expected to provide new insights in two major aspects of the physical world. First, in standard quantum mechanics, the  resolution-scale dependance is explicit in the Heisenberg uncertainty relations and several authors  \cite{feynmanhibbs1965,Nelson1966,abbotwise1981} commented on the fractal nature of the quantum path, which appears as a manifestation of the non-determinism of measurement outcomes. Second, complex and chaotic systems often involve the coupling between phenomena occurring at different scales and always demonstrate structures over broad ranges of scales. While chaos is a universal phenomenon investigated in many domains of science \cite{gleik1987,thompson2016}, there is no general framework for modeling these systems characterized by fractal dynamical paths with an at least effective loss of determinism.

The project of abandoning the hypothesis of differentiability seems daunting as differential calculus is the main mathematical tool of physics. We may however proceed without abandoning differential analysis tools by {\it smoothing out} any non-differentiable structures smaller than some parametric scale. In this paper we demonstrate this approach in the case of non-differentiable space coordinates while maintaining time as an external and absolute parameter. Furthermore, we are restricting ourselves to the type of non-differentiable paths with the resolution-scale dependence corresponding to quantum mechanical path or Brownian motion as motivated by Section \ref{fractal}.   Within this framework,  we do not proceed along in a standard relativistic approach, which would involve identifying invariant quantities and symmetries \cite{nottalecelerierlehner2006}. Instead of attempting to engage in such an ambitious program, we start by considering non-differentiable paths at a set resolution-scale and are lead to identify a doubling of the velocity field and to correspondingly define a complex resolution-scale dependent time-differential operator. This is done in Section \ref{differential}. This complex differential operator takes a familiar form and includes an additional higher order differential term, which is later shown to implement to the resolution-scale-covariance. We then continue by exploring the consequences of the use of this time differential operator in the usual development of a Lagrange mechanics with, however, a complex velocity and a complex Lagrange function as consequences of the non-differentiability of the dynamical paths. In Section \ref{lagrange}, applying a generalized stationary action principle, we show that the transition from the usual mechanics with differentiable paths to  a mechanics based on non-differentiable paths is simply implemented by replacing the usual time derivative with this new time-differential operator while keeping track of changes in the Leibniz product rule resulting from the higher order differential term. This effectively extends the principle of covariance to resolution scaling laws with the new time-differential operator playing the role of a {\it  resolution-scale-covariant derivative}. In Section \ref{harmonic}, we show that, under the restriction to stationary solutions, the fundamental relation of dynamics generalized to non-differentiable paths takes the form of a Langevin equation. We then proceed to the numerical integration of the Langevin equation in the case of a simple harmonic oscillator and show we recover the statistics of quantum mechanics for the same system. This is then formalized in Section \ref{schrodinger} where we see how the generalized fundamental relation of dynamics can be rewritten in the form of a Schr\"odinger equation. This motivates the interpretation of the system of axioms of quantum mechanics in terms of non-differentiable paths presented in Section \ref{axioms}. Finally, in Section \ref{chaos}, we then consider how the scale relativistic approach to quantum mechanics can be transposed to  complex or chaotic systems. This justifies the ongoing search for {\it quantum-like} signatures in the structures and dynamics of such systems. Section \ref{conclusion} then provides a summary and discussion.   

%%------------------------------------------------------------------------------------------------------------------------
%%------------------------------------------------------------------------------------------------------------------------
\section{Fractal dimension}\label{fractal}

We may approach the concept of fractality from the point of view of physical measurements. The measurement of any quantity $Q$ amounts to counting the number ${\mathcal M}_{\delta Q}$ of times  the unit quantity $\delta Q$ fits in $Q$.  The result of the measurement is then noted $Q={\mathcal M}_{\delta Q} \cdot \delta Q$. The measurement unit $\delta Q$ is generally chosen in a way that closely relates to the precision with which the measurement is carried out. For this reason and by simplification, we do not distinguish the resolution-scale from the measuring unit.   

In practice, it is usually implicitly assumed that two measurements of the same quantity performed with different resolution-scales ${\delta Q}$ and ${\delta Q'}$ are related by ${\mathcal M}_{\delta Q}\cdot \delta Q\approx{\mathcal M}_{\delta Q'}\cdot \delta Q'$ up to the combined experimental errors. The presumption then is that the quantity $Q$ does not depend on the scale of inspection. This leads to the idea that the measurement accuracy is {\it improved} by the use of a measuring device with an {\it improved} resolution.

This logic however breaks down when the structures contributing to the measurement outcome themselves depend on the scale of inspection. This is better seen when the quantity $Q$ is of geometrical nature, such as a length, an area or a volume. One may then write the resolution as $\delta Q=\left(\delta x\right)^{D_T}$ where $\delta x$ is the length scale with which the object is inspected (for simplification, we assume this resolution-scale to be the same in all directions, which is not necessarily the case) and $D_T$ is the {\it topological dimension} of the quantity being measured. 
Comparing measurements carried out with different resolution-scales, one may write $\left({\mathcal M}_{\delta x}/{\mathcal M}_{\delta x'}\right)=\left(\delta x/\delta x'\right)^{-D_F}$ with the fractal dimension $D_F=D_T+\delta D$, where $\delta D$, an anomalous exponent, is introduced to account for the resolution-scale dependance of the measurement. With this, $Q_{\delta x}/Q_{\delta x'}=\left({\delta x}/{\delta x'}\right)^{-D_F}\left(\delta x/\delta x'\right)^{D_T}=\left(\delta x/\delta x'\right)^{D_T-D_F}=\left(\delta x/\delta x'\right)^{-\delta D}$. This can be inverted to establish the fractal dimention of an object as $D_F=D_T- \frac{\log Q_{\delta x}/Q_{\delta x'}}{\log \delta x/\delta x'}$.   
\begin{figure}
% Use the relevant command to insert your figure file.
% For example, with the graphicx package use
 \includegraphics[width=0.60\textwidth]{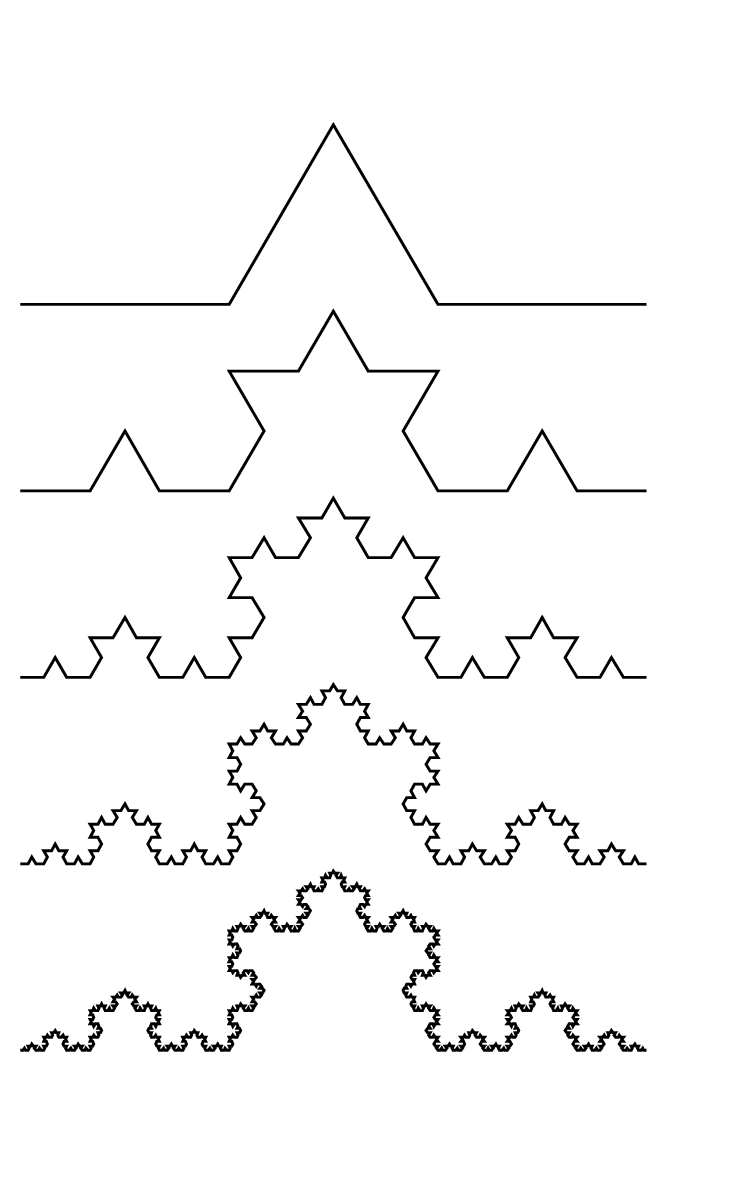}
  \includegraphics[width=0.40\textwidth]{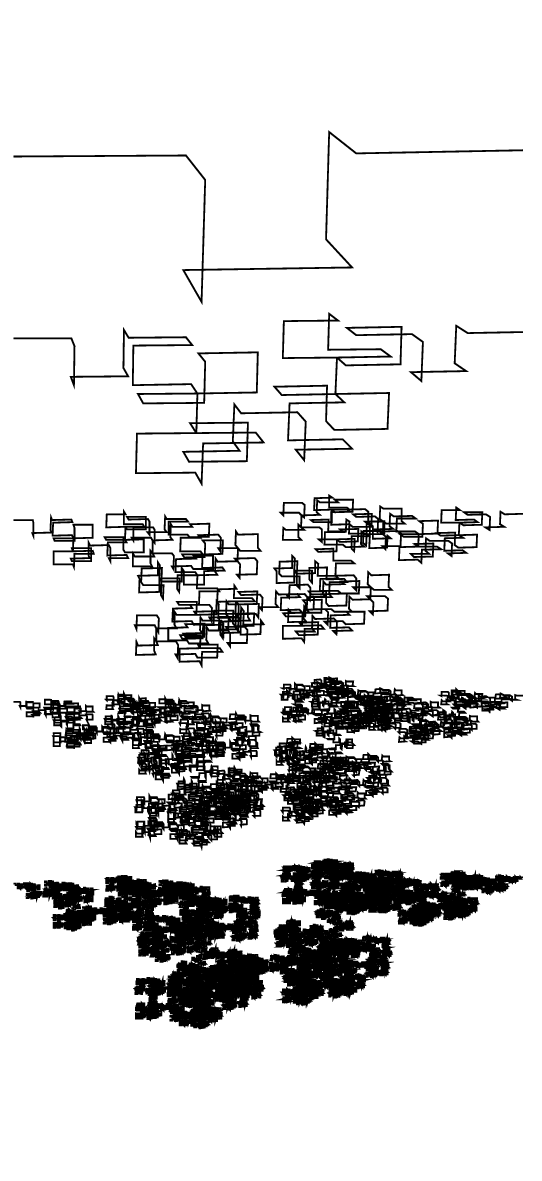}
% figure caption is below the figure
\caption{Left panel: Five iterations in the construction of the von Koch curve. Each iteration consists in replacing each segment of the previous iteration by 4 segments of length 3 times smaller, ultimately resulting in a curve of length diverging with resolution-scale at a rate characterized by the fractal dimension $D_F=\frac{\ln 4}{\ln 3}$. Alternatively, the successive curves can be regarded as representations of a same object inspected with different resolution-scales. Right panel: Cavalier projection of five iterations in the construction of a curve in three dimensions\cite{nottale1993}. Each segment of the previous iteration is replaced by 9 segments of length 3 times smaller, resulting in a fractal dimension $D_F=\frac{\ln 9}{\ln 3}=2$.}
\label{fractalfig}       % Give a unique label
\end{figure}
We may apply this to the case of the measurement of the length, $D_T=1$, of the von Koch curve (See Figure \ref{fractalfig}). Take $\delta x/\delta x'=3^p$, with $p$ some integer. Then by the construction of the von Koch curve, ${\mathcal M}_{\delta x}/{\mathcal M}_{\delta x'}=4^{-p}$ so that $Q_{\delta x}/Q_{\delta x'}=\left(3/4\right)^p$ and, $D_F=\log 4/\log 3\approx 1.261859\cdots$. The fact that $D_F>D_T$ indicates the divergence of the length as the curve is inspected at smaller scales. The right panel of Figure \ref{fractalfig} presents another example with a fractal dimension $D_F=2$. In self-similar objects such as the curves in Figure \ref{fractalfig}, the fractal dimension remains constant under successive changes of  resolution-scale. This, however, is not necessarily the case, and curves with a fractal dimension that depends on the resolution-scale are easy to imagine. We can even imagine changes of  resolution-scale resulting in transitions between fractal and non-fractal regimes. A very good example of this is the Brownian motion of a particle. The inspection of  the trajectory with a fine enough resolution reveals the kinks resulting from collisions with individual molecules of the surrounding fluid. Between collisions, the particle is observed in the ballistic regime\cite{mo2015}, with a path of fractal dimension 1.0, while at poorer resolution, in the diffusive regime, we will see below that the fractal dimension of the path is 2.0. Then, with an even coarser resolution, a drift may be revealed and dominate the motion with a path of fractal dimension of 1.0 again. 

In order to establish the fractal dimension of a Brownian path in the diffusive regime, it is useful to further concentrate on the specific case of the measurement of the length $\mathcal L$ of a curve. From what precedes, when a path is inspected with two different resolutions $\delta x$ and $\delta x'$, we have ${\mathcal L}_{\delta x}/{\mathcal L}_{\delta x'}=\left({\delta x}/{\delta x'}\right)^{1-D_F}$. Instead of measuring the length of the curve by inspection at a given resolution $\delta x$ or $\delta x'$, we may proceed by inspecting it at regular time intervals $\delta t $ or $\delta t'$. For this, we can consider the path over a time interval $T=\delta t \left({\mathcal L}_{\delta x}/\delta x\right)=\delta t' \left({\mathcal L}_{\delta x'}/\delta x'\right)$, in such a way $\delta t $ and $\delta t'$ are the times respectively required to travel distances $\delta x$ and $\delta x'$ on average . This implies $\delta t/\delta t'=\left(\delta x/\delta x'\right)\left({\mathcal L}_{\delta x'}/{\mathcal L}_{\delta x}\right)=\left(\delta x/\delta x'\right)^{D_F}$ and  ${\mathcal L}_{\delta t}/{\mathcal L}_{\delta t'}=\left(\delta t/\delta t'\right)^{\frac{1}{D_F}-1}$.

With this, we can consider the case of Brownian motion observed at resolution-scales corresponding to the diffusion regime. The distance traveled during a given time interval $\delta t$ scales with $\sqrt{\delta t}$ and the total distance traveled over the full duration $T$ of the observation is proportional to $\sqrt{\delta t}\, T/\delta t= T\left(\delta t\right)^{-1/2}$, corresponding to $D_F=2$, which implies a divergence of the velocity at small resolution-scale until the ballistic regime is reached. 

We can also consider the case of the path of a quantum particle. In 1965, Feynman and Hibbs wrote "{\it It appears that quantum-mechanical paths are very irregular}" and "{\it ($\cdots$) the 'mean' square value of a velocity averaged over a short time interval is finite, but its value becomes larger as the interval becomes shorter}" in such a way that "{\it $\cdots$ although a mean square velocity can be defined, no mean square velocity exists at any point. In other words, the paths are non-differentiable}"\cite{feynmanhibbs1965}. It should be noted that this preceded the word {\it fractal} coined by Beno\^it Mandelbrot in 1975 \cite{Mandelbrot1977}. In order to see this quantitatively from a simple argument \cite{abbotwise1981} based on the Heisenberg uncertainty relations, consider a particle of mass $m$ whose position is measured  $N$ times at regular time intervals $\Delta t$ over a time interval $T=N\Delta t$. If the particle is at rest on average, the measured path it travels results from the quantum fluctuations $\delta p$ of the momentum so that after $N$ measurements, the length of path that will have been recorded on average will be $\langle{\mathcal L}\rangle=N\frac{\delta p}{m}\Delta t$ or $\langle{\mathcal L}\rangle=\frac{\delta p}{m}T$. Using the Heisenberg uncertainty relation $\delta p\cdot \delta x \approx \hbar/2$, this may be rewritten $\langle{\mathcal L}\rangle=\frac{\hbar}{2m}\frac{T}{\delta x}$ and identifying with $\langle{\mathcal L}\rangle \propto \left(\delta x\right)^{1-D_F}$ we see that $D_F=2$. From one measurement to the next, the smallest significant change in path length $\langle\Delta {\mathcal L}\rangle$ is the position measurement resolution itself $\langle\Delta {\mathcal L}\rangle\approx \delta x$. With this, we obtain $\delta x^2\approx \frac{\hbar}{2m}\Delta t$. Comparing this with $\delta t\propto \delta x^{D_F}$ we identify the effective diffusion coefficient ${\mathcal D}=\frac{\hbar}{2m}$. 

In the following section, we establish a general method of approach to these non-differentiable and resolution-scale dependent paths and we apply this method specifically to the case $D_F=2$ because of it's importance in physics.

%%------------------------------------------------------------------------------------------------------------------------
%%------------------------------------------------------------------------------------------------------------------------
\section {Scale covariant time-differential operator}\label{differential}
\begin{figure}
\includegraphics[width=0.95\textwidth]{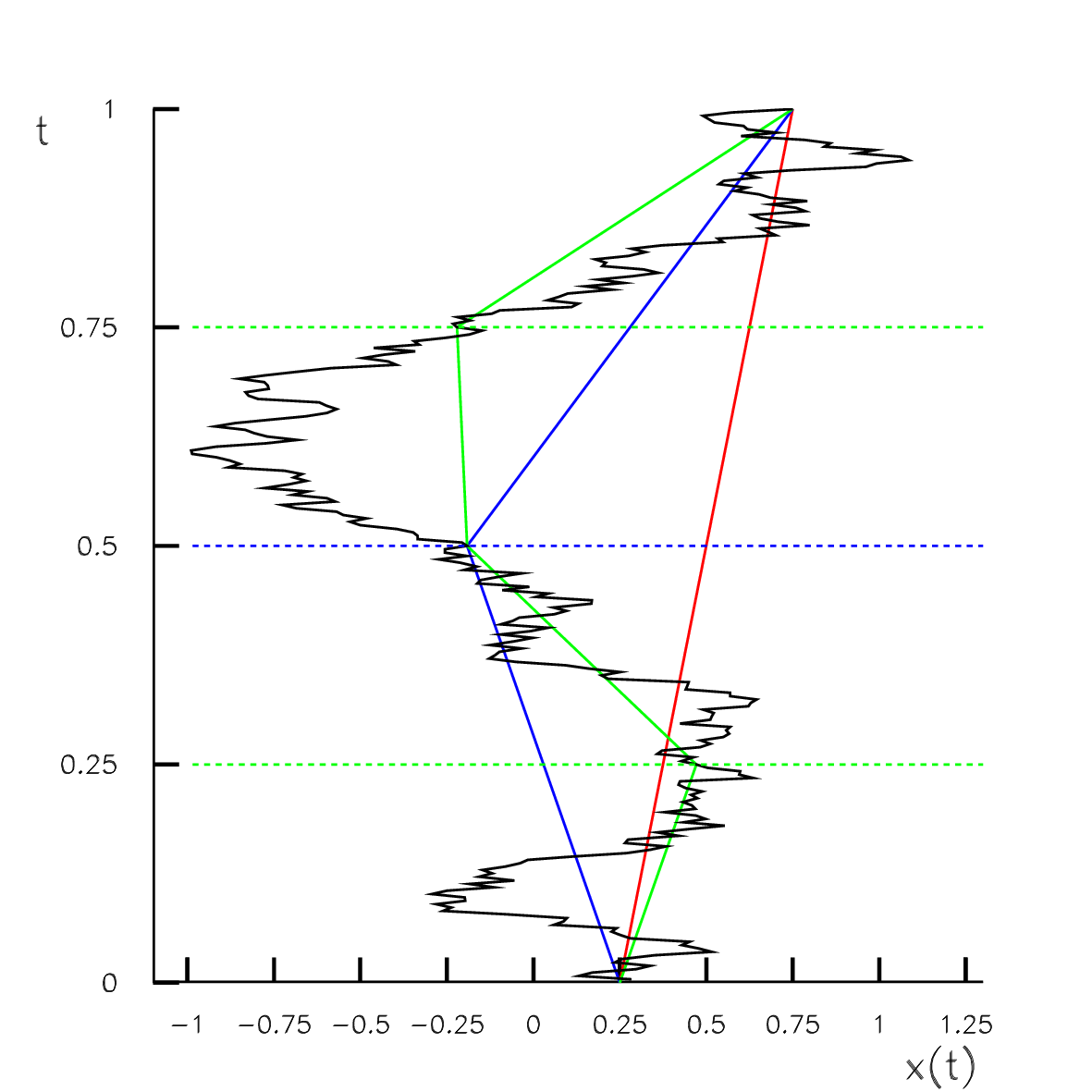}
\caption{The black curve represents an approximation $x(t)$ of a non-differentiable path $q(t)$. It was obtained by dichotomy iterations,  each refining the resolution-scale $\delta t$ by a factor two. The red line represents the first iteration, the blue line the second, the green line the third and the black line the ninth. When going from one iteration to the next, the position $x$ of the midpoint of each segment is shifted randomly in such a way that, on average, the combined displacements corresponding to the two resulting segments is $\sqrt{2}$ larger than the displacement represented by the original one. This corresponds to a fractal dimension of 2.0. Different iterations can be regarded as representations with different time intervals of inspection  $\delta t$ of a nowhere differentiable curve.  The slope of the segments of a broken line corresponding to a specific resolution-scale are related to the usual velocities ${\bf v}_\pm(\delta t)$ before and after each sampled point. The residual between a broken line and the curve is a stochastic process ${\bf b}_\pm$.  As the curve is considered with finer resolutions, the before and after finite differences in $t=0.5$ for example keep fluctuating with an amplitude that diverges. }
\label{fractcrv}
\end{figure}
The usual derivative $f'(t)$ of a function $f(t)$ is defined by considering the change of the value of the function in the limit of infinitesimally small increase $(+)$ or decrease $(-)$ of the argument:
$$
f'_+(t)=\lim_{\delta t\to 0^+}\frac{f(t+\delta t)-f(t)}{\delta t}{\rm ~~and~~ }f'_-(t)=\lim_{\delta t\to 0^+}\frac{f(t)-f(t-\delta t)}{\delta t}
$$
with $f'(t)=f_+'(t)=f_-'(t)$ remaining single valued as long as the function is differentiable in $t$. If, on the contrary, the function is continuous but non-differentiable in $t$, then $f'(t)$ is no longer uniquely defined as $f_+'(t)\neq f_-'(t)$. Furthermore, if the function is non-differentiable in a dense
 set of points (a set  is said dense if any neighborhood of any point not in the set contains at least one point in the set) , then, taking the limit $\delta t\to 0^+$ becomes impossible in practice as the outcome fluctuates indefinitely with a diverging amplitude as the limit is approached (see Figure \ref{fractcrv}). We then have to refrain from taking the limit and consider a resolution-scale dependent function $f(t,\delta t)$, which is a {\it smoothed out version} of the original function $f(t)$.  This then leads to defining a double-valued and  explicitly resolution-scale dependent differential: 
\begin{eqnarray*}
f'_+(t,\delta t)&=&\frac{f(t+\delta t,\delta t)-f(t,\delta t)}{\delta t}{~~~~\delta t>0};\\
f'_-(t,\delta t)&=&\frac{f(t+\delta t,\delta t)-f(t,\delta t)}{\delta t}{~~~~\delta t<0}.
\end{eqnarray*}
With this in mind, we can consider the displacement of a point of position ${\bf q}(t)$ along a non-differentiable path with the following representation with two terms: 
\begin{eqnarray}\label{repres}
d{\bf q}_+&=&{\bf v}_+dt+d{\bf b}_+{~~~~0<dt<\delta t}\nonumber \\
d{\bf q}_-&=&{\bf v}_-dt+d{\bf b}_-{~~~~\delta t<dt<0}
\end{eqnarray}

The first term proceeds from {\it usual velocities} ${\bf v}_\pm$ before and after time $t$, and depends on the scale of inspection $|\delta t|$ (see Figure \ref{fractcrv}). By {\it usual velocity} we mean a finite displacement  $\delta{\bf x}$ divided by the time $\delta t$ it takes to complete it.  The displacement $\delta{\bf x}$ or the time $\delta t$ correspond to the resolution-scale with which the path is inspected to yield a specific representation. 

On the contrary, ${d\bf b}_\pm(t)$ represents the residual, possibly a stochastic process, to be revealed by finer observations. There is an infinite number of paths whose inspection at a finite resolution $\delta t$ yield the same representation. They differ from each other only by their respective $d{\bf b}_\pm$.  Consequently, the expectation value of $d{\bf b}_\pm$ over this ensemble must cancel $\langle {d\bf b}_\pm\rangle=0$. Indeed, the non-cancelation of this expectation value would betray some knowledge about the path inspected with a resolution finer than actually considered. We will shortly come back to the statistical properties of ${d\bf b}_\pm(t)$ as they determine how the path representation is affected by a change in resolution-scale. Indeed, the {\it usual velocity} at a given resolution $\delta t$ can be seen as deriving from the residual ${d\bf b}_\pm(t)$ in the representation of the same path inspected at a coarser resolution. 

One could come up with multitude of alternative resolution-scale specific representations of the path, each corresponding to a different smoothing out of details smaller than the inspection scale and to be regarded as a stochastic residual. They should all be equivalent up to the considered resolution-scale and, while we do not provide any formal proof of this, different representations are not expected to affect the subsequent developments. Indeed, it is worth stressing that this two-term representation as well as any other such possible representation of the path  is in fact a formalization of the usual way we deal with trajectories:  details smaller than some tractable or practical resolution-scale are disregarded or smoothed out. In particular, the infinitesimal limit $\delta t\to 0$ is never actually taken in a strict sense as it would in fact incapacitate the measurment process. The step being taken here can be regarded as the promotion of the resolution-scale to be one of the relative characteristics of reference frames, at the same level as position, orientation and motion in a way which constitutes the essence  of the scale relativity approach\cite{nottale1993,nottale2011}.  

With this, concentrating on one specific resolution-scale $\delta t$, we average out the stochastic residuals ${d\bf b}_\pm(t)$ by defining explicitly resolution scale dependent {\it classical} time-differentials as the expectation values of the time-differentials {\it after} and {\it before} the considered point:
$$
\dot{\bf x}_+=\frac{d_+}{dt}\bigg |_{\delta t}{\bf q}={\bf v}_+ +\langle\frac{d{\bf b}_+}{dt}\rangle={\bf v}_+{\rm ~~and~~} \dot{\bf x}_-=\frac{d_-}{dt}\bigg |_{\delta t}{\bf q}={\bf v}_-+\langle\frac{d{\bf b}_-}{dt}\rangle={\bf v}_-
$$
Furthermore,  rather than continuing to manipulate the two {\it classical } differential operators separately, we combine them linearly into a single {\it complex time-differential operator}\,\cite{nottale1993}: 

\begin{eqnarray}\label{complex_diff_op}
\frac{\hat d}{dt}=\frac{1}{2}\left(\frac{d_+}{dt}\bigg |_{\delta t}+\frac{d_-}{dt}\bigg |_{\delta t}\right)-\frac{i}{2}\left(\frac{d_+}{dt}\bigg |_{\delta t}-\frac{d_-}{dt}\bigg |_{\delta t}\right).
\end{eqnarray}
The real part corresponds to the average of the {\it after} and {\it before} finite $\delta t$-differentials. It can be thought of as the classical differential, which is preserved for a differentiable function in the limit $\delta t\to 0$. The imaginary part is the halved difference between the {\it after} and {\it before} finite $\delta t$-differentials. It can be thought of a the {\it kink differential} which vanishes  for a differentiable function in the limit $dt\to 0$. The choice of the sign of the imaginary part is of no consequence. It is made in such a way that the following development of dynamics equations leads to an identification of wave functions rather than their complex conjugates.

When acting on $\bf x$ with the  complex time-differential operator, we can define the {\it complex velocity}:
\begin{eqnarray}
\label{complexv}
\mathcal V=\frac{\hat d}{dt}{\bf q}=\frac{{\bf v}_++{\bf v}_-}{2}-i\frac{{\bf v}_+-{\bf v}_-}{2}={\bf V}-i{\bf U}
\end{eqnarray}
where $\bf V$ can be regarded as the classical velocity and $\bf U$ is an additional term, the {\it kink velocity} which persists under the inspection of non-differentiable paths with ever finer resolutions.

Equipped with these definitions, we can consider a regular differentiable field $h({\bf x},t)$, and write its total derivative as a Taylor expansion (the repetition of an index means the implicit summation over that index):
$$
\frac{dh}{dt}=\frac{\partial h}{\partial t}+\frac{\partial h}{\partial x_i}\frac{dx_i}{dt}+\frac{1}{2}\frac{\partial^2 h}{\partial x_i\partial x_j}\frac{dx_idx_j}{dt}+\frac{1}{6}\frac{\partial^3 h}{\partial x_i\partial x_j\partial x_k}\frac{dx_idx_jdx_k}{dt}+\cdots
$$

We may consider $h({\bf q},t)$ along a non-differentiable path ${\bf q}(t,\delta t)$ in a two-term finite resolution-scale representation (Equation \ref{repres}). As we already did in the definition of $\frac {\hat d}{dt}$, we proceed at a set resolution-scale and average out the residual stochastic part to be revealed only at finer resolution-scales. Consequently, terms of sufficiently high order must be included so as to give provision for all non-vanishing contributions from the residual components $d\bf b_\pm$, which may be written as fractional exponents of the time element $dt$.

At this point we depart from generality and  make a choice as to the statistical nature of the residual process ${\bf b}_\pm(t)$. We would like to concentrate on the case $D_F=2$, specific to Brownian motion and quantum mechanical paths (See Section \ref{fractal}). We then consider $d\bf b_\pm$  as a Wiener process with  $\langle d{\bf b}_\pm\rangle=0$ as already discussed, $\langle db_{i+}\cdot db_{i-}\rangle=0$,  and $\langle db_{i+}\cdot db_{j+}\rangle=\langle db_{i-}\cdot db_{j-}\rangle=2\mathcal D\delta_{i,j}|dt|$ with $\mathcal D$ akin to a diffusion coefficient. It should be stressed that while this choice corresponds to the very general case of Markovian or random walks to which we are restricting ourselves here, infinitely many other forms could be explored for the description of stochastic process that are not memoryless. The underlying statistical distribution can naturally be thought of as Gaussian.  However, by virtue of the central limit theorem, any other statistical distribution would be equally valid and would make no difference for the rest of the development. 
So, just as we did before, we now consider both the {\it after} and the {\it before} time-differentials while keeping only the terms that do not vanish with $dt$. Note that in taking the limit here, we do not change the resolution-scale $\delta t$ as we are now considering a specific representation of the path with a classical component linear in $dt$ and a stochastic component accounted for through an expectation value with a fractional power $1/2$ of $|dt|$: 
$$
\frac{d_\pm }{dt}h=\frac{\partial h}{\partial t}+\frac{\partial h}{\partial x_i}v_{i\pm}+\frac{\partial h}{\partial x_i}\langle\frac{db_{i\pm}}{dt}\rangle+\frac{1}{2}\frac{\partial^2 h}{\partial x_i\partial x_j}\langle\frac{db_{i\pm}db_{j\pm}}{dt}\rangle
$$
Since ${d\bf b_\pm}$ is of order $\sqrt{|dt|}$, the third term would diverge with $dt\to 0$ if it were not for the expectation value, which makes it cancel as $\langle d{\bf b}_\pm\rangle=0$. The {\it after} and {\it before} time-differentials become:
$$
\frac{d_\pm}{dt}h=\frac{\partial h}{\partial t}+{\bf v}_{\pm}\cdot\nabla h \pm \mathcal D\nabla^2 h
$$
Combining the {\it after} and {\it before} differential operators in a single complex differential operator as before:
\begin{eqnarray}\label{covardiff}
\frac{\hat d}{dt}=\frac{\partial }{\partial t}+\mathcal V\cdot \nabla -i \mathcal D\Delta 
\end{eqnarray}
It can be anticipated that going from differentiable geometry to non-differentiable geometry should be implemented by replacing the usual time derivative $\frac {d}{dt}$ with this complex time-differential operator $\frac{\hat d}{dt}$ provided proper attention is given to the changes in the Leibniz product rule implied by the second derivative appearing in the last term of Equation \ref{covardiff}. In the next section, we verify that this is indeed the case for the Lagrange formulation of mechanics. 
%%------------------------------------------------------------------------------------------------------------------------
%%------------------------------------------------------------------------------------------------------------------------
\section{Mechanics of non-differentiable paths}\label{lagrange}
Considering non-differentiable paths with the corresponding double-valued velocities with  {\it after} and {\it before} components $\dot {\bf x}_\pm$, we assume here that the mechanical system with the configuration coordinate ${\bf x}$ can be characterized by a now complex Lagrange function $\mathcal L({\bf x},\mathcal V,t)$. We can then express the action for the evolution of the system between times $t_1$ and $t_2$ as $\mathcal S=\int_{t_1}^{t_2}\mathcal L({\bf x},\mathcal V,t)dt$. As we are about to proceed with the optimization of the action, it is worth stressing here that neither $\mathcal V$, $\mathcal L$ nor $\mathcal S$ are stochastic. As an argument of the Lagrange function, the position $\bf x$ is a parameter independent of time. In the calculation of the action $\mathcal S$, the position $\bf x$ is a smoothed out version of the possibly stochastic path $\bf q$ with all the features at time scales less than some resolution time scale $\delta t$ averaged out. The complex velocity $\mathcal V$ results from the action on $\bf q$ of $\frac{\hat d}{dt}$, which averages out the stochastic components over the inspection time-scale $\delta t$. It embodies the velocity doubling resulting from the non-differentiable nature of the path observed at a specific resolution scale bellow which other details are smoothed out. Hence, all these functions are defined in a way that depends on the inspection time scale $\delta t$  playing the role of the resolution-scale. So, while the following development present similarities with what is done in stochastic mechanics \cite{Yasue1981,Koide2015}, the motivations and the objects manipulated here are quite different.

In order to lighten the notation without loosing any generality, we proceed by considering a one-dimensional problem with  $\mathcal V=\frac{1}{2}\left(\dot x_++\dot x_-\right)-\frac{i}{2}\left(\dot x_+-\dot x_-\right)$ and $\mathcal L(x,\mathcal V,t)=\mathcal L(x,\frac{1-i}{2}\dot x_++\frac{1+i}{2}\dot x_-,t)$.  The  generalized stationary action principle can be written as
$$
\delta\mathcal S=\int_{t_1}^{t_2}\left(\frac{\partial \mathcal L}{\partial x}\delta x+\frac{\partial \mathcal L}{\partial \dot x_+}\delta \dot x_++\frac{\partial \mathcal L}{\partial \dot x_-}\delta \dot x_-\right)dt=0.
$$   

Provided the complex Lagrange function is an analytic function of $\mathcal V$ and making use of $\frac{\hat d}{dt}$ (See Equation \ref{complex_diff_op}), it becomes:
\begin{eqnarray}
\delta\mathcal S&=&\int_{t_1}^{t_2}\left(\frac{\partial \mathcal L}{\partial x}\delta x+\frac{\partial \mathcal L}{\partial \mathcal V}\frac{1-i}{2}\delta \dot x_++\frac{\partial \mathcal L}{\partial \mathcal V}\frac{1+i}{2}\delta \dot x_-\right)dt\nonumber\\
&=&\int_{t_1}^{t_2}\left(\frac{\partial \mathcal L}{\partial x}\delta x+\frac{\partial \mathcal L}{\partial \mathcal V}\frac{\hat d}{dt}\left(\delta x\right) \right)dt=0\nonumber
\end{eqnarray}
%$$
%\delta\mathcal S=\int_{t_1}^{t_2}\left(\frac{\partial \mathcal L}{\partial x}\delta x+\frac{\partial \mathcal L}{\partial \mathcal V}\frac{1-i}{2}\delta \dot x_++\frac{\partial \mathcal L}{\partial \mathcal V}\frac{1+i}{2}\delta \dot x_-\right)dt=\int_{t_1}^{t_2}\left(\frac{\partial \mathcal L}{\partial x}\delta x+\frac{\partial \mathcal L}{\partial \mathcal V}\frac{\hat d}{dt}\left(\delta x\right) \right)dt=0
%$$   
We recover the usual form of the expression of action stationarity with the velocity replaced with the complex velocity $\mathcal V$ and the time derivative replaced with the complex time-differential $\frac{\hat d}{dt}$.  From this point, we need to integrate by parts. This requires care as $\frac{\hat d}{dt}$ includes a second derivative which affects the product rule. One can verify that $\frac{\hat d}{dt}\left(f\cdot g\right)=\frac{\hat d f}{dt}g+f\frac{\hat d g}{dt}-2i\mathcal D\nabla f\cdot \nabla g$ and we obtain 
$$
\delta\mathcal S=\int_{t_1}^{t_2}\left(\frac{\partial \mathcal L}{\partial x}\delta x+ \frac{\hat d}{dt}\left(\frac{\partial \mathcal L}{\partial \mathcal V}\cdot \delta x\right)-\frac{\hat d}{dt}\left(\frac{\partial \mathcal L}{\partial \mathcal V}\right)\delta x\right)dt=0,
$$
where the term in $\frac{ \partial \delta x}{\partial x}$ was discarded as $\delta x$ is not a function of $x$. Considering that $\delta x(t_1)=\delta x(t_2)=0$ and requiring  this equation to hold for any infinitesimal $\delta x(t)$, we obtain the usual Euler-Lagrange equation but with the complex velocity and time-differential operator: 
$$
\frac{\partial \mathcal L}{\partial x}-\frac{\hat d}{dt}\left(\frac{\partial \mathcal L}{\partial \mathcal V}\right)=0
$$

Using the usual form of the kinetic energy and  including a purely real potential energy term $\Phi$ associated with a conservative force acting on the particle : $\mathcal L=\frac{1}{2}m\mathcal V^2-\Phi$, the Euler-Lagrange equation results in a generalized form of Newton's relation of dynamics
\begin{eqnarray}\label{newton}
m\frac{\hat d}{dt}\mathcal V=-\nabla\Phi.
\end{eqnarray}

The recovery of  the velocity and time derivative replaced with their complex conterparts $\mathcal V$ and ${{\hat d}\over{dt}}$ indicates that this replacement implements the transition from the usually assumed differentiable geometry to a non-differentiable geometry with ${{\hat d}\over{dt}}$ playing the role of a {\it scale-covariant derivative}. In the next section, we explore the implications of this transition in the case of one of the simplest mechanical systems: the harmonic oscillator.  
%%------------------------------------------------------------------------------------------------------------------------
%%------------------------------------------------------------------------------------------------------------------------
\section{Application to the one-dimensional harmonic oscillator}\label{harmonic}
The fundamental relation of dynamics obtained above (Equation \ref{newton}) has both real and imaginary parts, which we can write separately, replacing $\mathcal V$  and $\frac{\hat d}{dt}$ by their expressions (Equations  \ref{complexv} and \ref{covardiff}). 
\begin{eqnarray}
\frac{\partial}{\partial t}{\bf V}-\mathcal D\Delta{\bf U}+({\bf V}\cdot\nabla){\bf V}-({\bf U}\cdot\nabla){\bf U}&=&-\frac{1}{m}\nabla\Phi\nonumber\\
\frac{\partial}{\partial t}{\bf U}+\mathcal D\Delta{\bf V}+({\bf V}\cdot\nabla){\bf U}+({\bf U}\cdot\nabla){\bf V}&=&0\nonumber
\end{eqnarray}

This system of differential equations is the same as Equations (34) in E. Nelson's article entitled "Derivation of the Schr\"odinger equation from Newtonian Mechanics" \cite{Nelson1966}. However, they originate quite differently. In Nelson's stochastic quantization, these equations result from the additional hypothesis of some underlying Brownian motion, which, in the quantum mechanical context, is characterized by a diffusion coefficient $\mathcal D=\frac{\hbar}{2m}$. In the scale relativity approach followed here, these equations result from the abandonment of the usually implicit differentiability hypothesis with the choice $\langle db_{i\pm}\cdot db_{j\pm}\rangle=2\mathcal D\delta_{i,j}dt$. 

We may consider differentiable and classical solutions for which the {\it kink velocity} ${\bf U}=0$. The first equation then appears as the usual fundamental relation of dynamics $\frac{d{\bf V}}{dt}=-\frac{1}{m}\nabla{\Phi}$ and the second becomes $\Delta {\bf V}=0$ which is already ensured by the first one as $\bf V$ no longer can be an explicit function of position ${\bf x}$.

Inversely, we may concentrate on a new type of stationary motion characterized by $\langle{\bf V}\rangle=0$. It emerges entirely as a consequence of the non-differentiability. Among all the possible paths with this property, in order to simplify the above system of differential equations, we may choose a subset  such that ${\bf V}=0$ : 
\begin{eqnarray}
\mathcal D\Delta{\bf U}+({\bf U}\cdot\nabla){\bf U}&=&\frac{1}{m}\nabla\Phi\nonumber\\
\frac{\partial}{\partial t}{\bf U}&=&0\nonumber
\end{eqnarray}
The second equation implies that $\bf U$ depends only on $\bf x$. The solution $\bf{U}({\bf x})$ to the first equation can be used in a time forward Langevin equation. Indeed, since we chose ${\bf V}=0$, we have ${{\bf v}_+=-{\bf v}_-}$, and ${\bf U}={\bf v}_+$. Consequently, using Equation \ref{repres} with a finite time step $\delta t$ corresponding to the resolution-scale gives
\begin{equation}
\label{langevin}
\delta{\bf x}_+={\bf U}({\bf x})\delta t+\delta{\bf b}_+
\end{equation}
where $\delta{\bf b}_+$ is a stochastic term such that $\langle \delta{\bf b}_+ \rangle=0$ and $\langle \delta b_{i+}\cdot \delta b_{j+}\rangle=2\mathcal D\delta_{i,j}\delta t$.
Considering a finite time step, a natural choice for  $d{\bf b}_+$ is a Gaussian deviate  of zero mean and with a standard deviation $\sqrt{2\mathcal D \delta t}$.

McClendon and Rabitz \cite{McClendon1988} simulated several quantum systems using the differential equations of Nelson's stochastic quantization as a starting point \cite{Nelson1966}. The case of an infinite square well has been studied by Hermann \cite{Hermann1997} with the scale relativity approach presented here. Even more recently, the finite square well was also studied by Al-Rashid et al.\cite{alrashid2011}. Nottale \cite{nottale2011} also simulated Young one and two-slit experiments as well as the hydrogen atom. Here, we consider the case of a one-dimensional harmonic oscillator for which $\Phi(X)=\frac 12 m\Omega^2 X^2$, where $m$ is the mass of the particle and $\Omega$ is the frequency of the oscillator. In one dimension, the differential equation for $\bf U$  becomes: 
$\mathcal D\frac{d^2U}{dX^2}+U\frac{dU}{dX}=\frac{d}{dX}\left(\mathcal D\frac{dU}{dX}+\frac 12 U^2  \right)=\Omega^2 X$
and integrating once,  $\mathcal D\frac{dU}{dX}+\frac 12 U^2 +C=\frac 12 \Omega^2 X^2$ where $C$ is an integration constant with the dimension of the square of a velocity. Introducing 
the dimensionless variables $u=\frac{U}{\sqrt{2\mathcal D\Omega}}$, $x=\sqrt{\frac{\Omega}{2\mathcal D}}X$, and $c=\frac{C}{\mathcal D \Omega}$, this equation takes the form
$$
\frac{du}{dx}+u^2+c= x^2
$$ 

This is a first order non-linear differential equation of the Riccati form \cite{bonilla2017,saidalrashid} reducible to a linear second order differential equation for a function $y$,  
$$
\frac{d^2y}{dx^2}-x^2y=-cy,
$$
which is the differential equation satisfied by Hermite functions provided the constant $c$ equals $c_n=2n+1$ where $n$ is an integer. We may note that $E_n=mC_n=2m\mathcal D \Omega(n+\frac 12)$ correspond to the eigen-energies of the quantum harmonic oscillator for the identification $\hbar=2m\mathcal D$. The corresponding solutions are Hermite functions $y_n$ and each is associated to a solution of the Riccati equation as $u_n=\frac{1}{y_n}\frac{dy_n}{dx}$ or $u_n(x)=-x+\frac{1}{H_n}\frac{dH_n}{dx}$ with $H_n$ the Hermite polynomial of order $n$.

In the limit $|x|\to\infty$, the term $\frac{1}{H_n}\frac{dH_n}{dx}$ vanishes in comparison to $-x$. 
As a consequence, in the Langevin equation \ref{langevin}, and in the limit $|x|\to\infty$, the path {\it kinked inward}, toward the center of the well in proportion to the distance. Similarly, in Equation \ref{langevin}, the path is {\it kinked away} from the regions where the Hermite polynomials have a root. 

It is interesting to see how the solutions $u_n$ would change if we considered the integration constant $c$ to depart from $c_n$. We may write $c=c_n+\delta c$ and  $u=u_n+\xi$. The differential equation gives: $\frac{d\xi}{d x}+\xi^2+2 u_n \xi+\delta c=0$. Far from the well center, $|x|\gg1$ and $u_n= -x$ and we can consider $|\xi|\ll |u_n|$ so the $\xi^2$ term can be dropped. With this simplification and writing $\xi=f(x)e^{ x^2}$, it comes $\frac {df}{d x}=-\delta Ce^{- x^2}$ and $f( x)=-\frac{\delta c \sqrt{\pi}}{2}{\rm sign}(x)$ so that $ u( x)\approx- x -\frac{\delta c \sqrt{\pi}}{2}{\rm sign}(x)e^{ x^2}$. In the Langevin equation \ref{langevin}, the corrective term will be responsible for the path to be either kinked away from the center of the well or kinked toward the center of the well depending on the sign of $\delta c$. The path is either forced toward the center or escapes indefinitely, in both cases departing from the requirement that $\langle V\rangle=0$. This, in itself, is reminding of the fact that when solving the time independent Schr\"odinger equation for the harmonic well, the wave functions can be normalized only for the eigen-energies \cite{griffiths1995}. 

\begin{figure}
\includegraphics[width=0.45\textwidth]{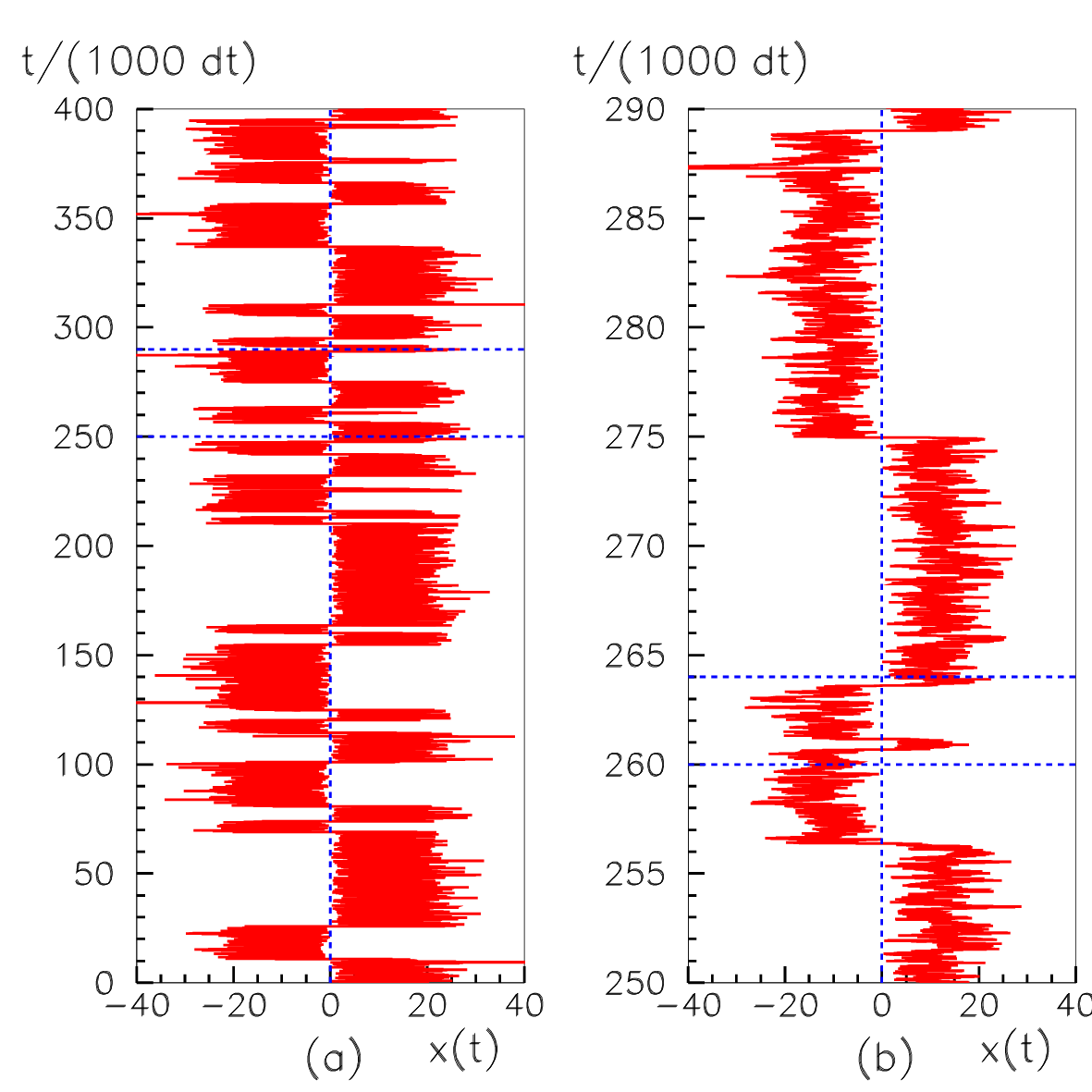}
\includegraphics[width=0.45\textwidth]{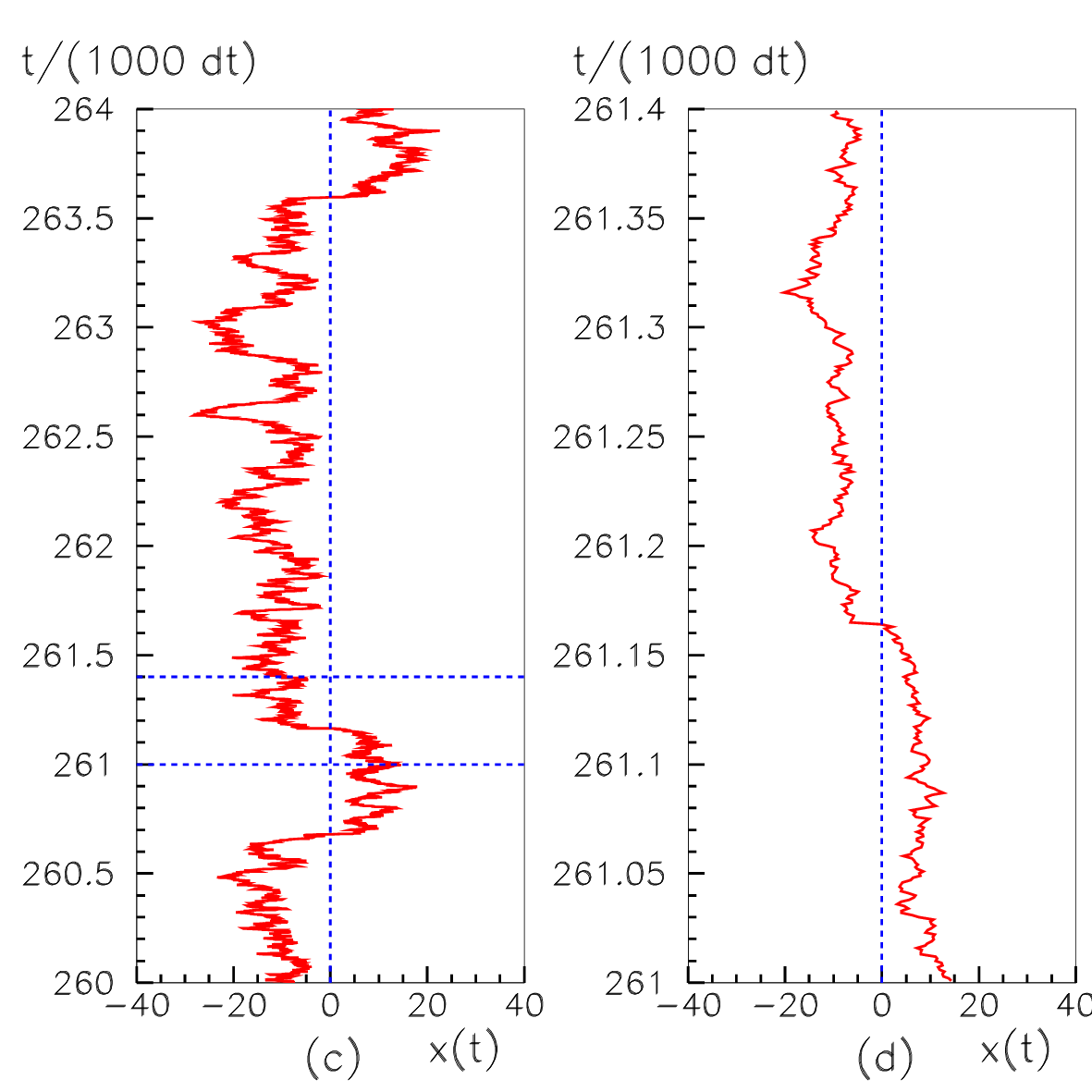}
\caption{The simulated path for a harmonic oscillator random walk with $\omega=0.01$  in mode $n=2$ is presented over $4\times 10^5$ time steps in the left most graph $(a)$.  Graph $(b)$ expands graph $(a)$ concentrating on the time interval delimited by the two horizontal dashed lines and so on and so forth for graphs $(c)$ and $(d)$, each time zooming in on a time interval ten times smaller. }
\label{path}
\end{figure}

We numerically integrated the Langevin equation with  $\Omega=0.001 \delta t^{-1}$ while using $\sqrt{2\mathcal D \delta t}$ as a distance unit.  Figure \ref{path} shows the simulated path in a harmonic oscillator in mode $n=2$. In panel $(a)$, the path appears to spend as much time on either side of the well and is rarely found close to the center of the well. The center of the well appears as a {\it node} separating two {\it lobes}.  The symmetry degrades as shorter time intervals are considered in panels $(b)$, $(c)$ and $(d)$. When shorter time intervals are considered, the path is increasingly likely to be found either in one {\it lobe} or another. When the path is in the region corresponding to one {\it lobe}, in order to migrate to a different {\it lobe}, it needs to undergo a large enough hop. Because of the Gauss distribution of the hops, large ones are very infrequent and the path spends a varying amount of time in each {\it lobe}.

\begin{figure}
\includegraphics[width=0.95\textwidth]{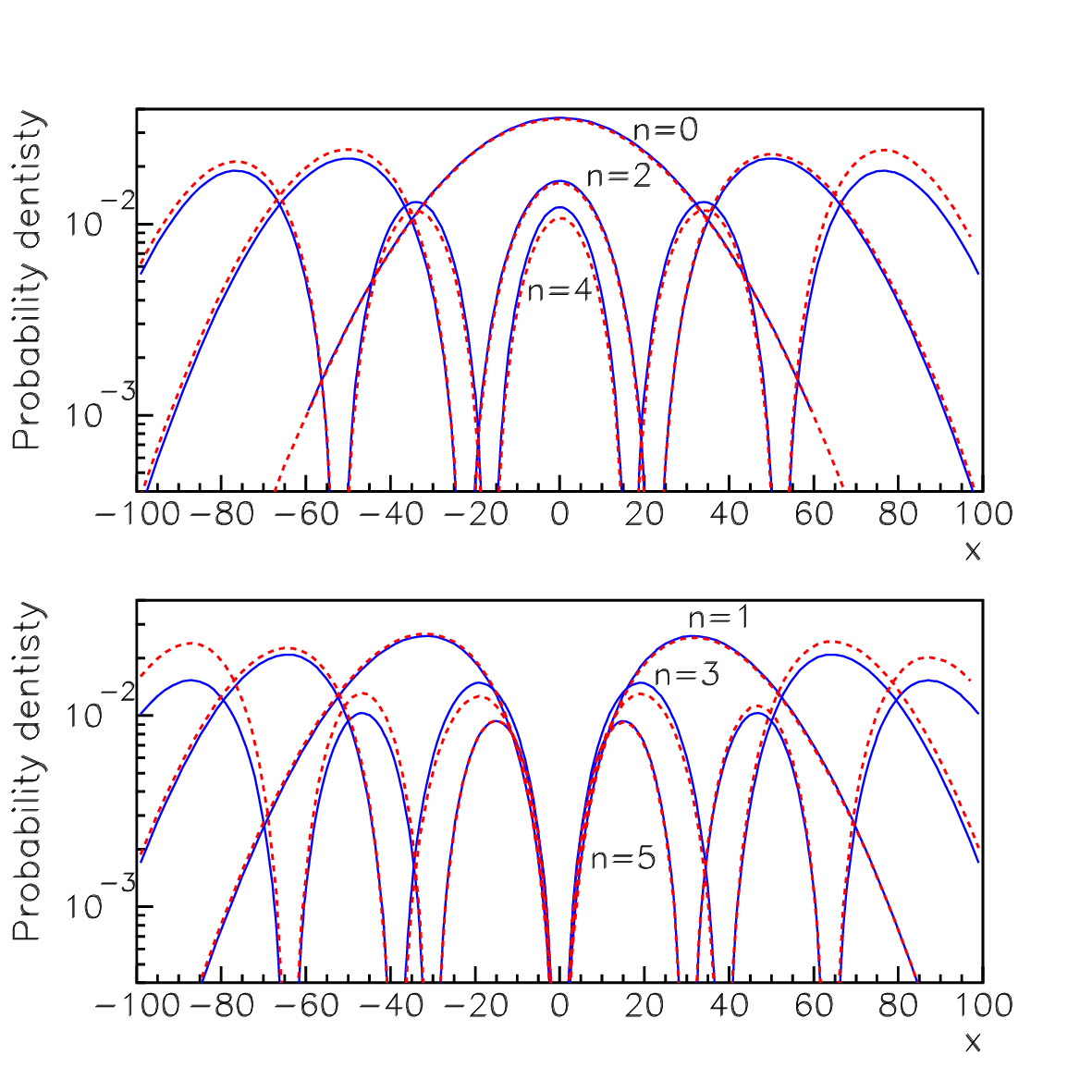}
\caption{The histograms of the position of the harmonic oscillator with $\omega=0.001$ followed in 20 numerical experiments of $10^7$ steps for n=0, 2 and 4 (top) and n=1, 3 and 5 (bottom) are shown by the red dashed curves. The solid blue curves represent the Hermite functions, solution of the Schr\"odinger equation for the same harmonic oscillator. }
\label{pdens} 
\end{figure}

Figure \ref{pdens} shows the histograms of the positions of the path for $n$ ranging from 0 to 5.  Even and odd values of $n$ were separated in the upper and lower panels respectively to improve visibility. For each value of $n$, 20 independent numerical experiments were performed over $10^7$ unit time steps $\delta t$, each starting from a randomly chosen point. The red curves represent the histograms, which are compared to the squares of Hermite functions, solutions of the Schr\"odinger equation, represented by the blue curves. The unevenness of the {\it lobes} of the red curves results from statistical fluctuations discussed above. The unevenness of the {\it lobes} is reduced as the duration of the simulation is increased. 

The fact that these simulations with $V=0$ reproduce the familiar solutions of the time independent Schr\"odinger equation is suggestive of quantum mechanics being a manifestation of the non-differentiability of the paths. This result depends on the very specific choice $\langle \delta b_{i\pm}\cdot d\delta b_{j\pm}\rangle=2\mathcal D\delta_{i,j}\delta t$ we have made for the statistical property the stochastic term in the path representation (See Equation \ref{repres}).  It should be stressed however that these paths, simulated with a finite time step $\delta t$, are nothing more than a sampling of geometrical points along a non-differentiable mechanical paths in an infinite set. Only the infinite set of paths may be identified to the state evolution of the quantum particle. One simulated path should not be regarded in any way as an actual trajectory followed by a quantum particle. 

Indeed, let us imagine that Figure \ref{path} represents the path of a physical particle, which, in-between time steps, travels along a differentiable trajectory, in such a way that any finer resolution would not reveal any new structures in the path. Then, from panel $(a)$ to panel $(d)$ in Figure \ref{path}, we observe the progressive transition from a {\it fractal} path at large time scales toward a {\it differentiable} path at shorter time scale. If the simulation had been carried out for the same time duration but with a time step $1000$ times smaller, then the right most panel (d) would look like the leftmost panel (a). Meanwhile, in panel $(a)$, the number of hopping from one side to the other, would be so great that they would be indistinguishable and for all practical purposes in that graph, at any given time, the particle could  only be probabilistically described as being on one {\it lobe} or the other. Pushing this even further by making the elementary time step tend to zero, all four panels of Figure \ref{path} would have the exact same appearance as we would be infinitely far away from the scale at which the transition from {\it probabilistic} behavior to {\it trajectory-like} behavior may take place. Even though we have been thinking about paths in the usual sense, in the limit of infinitesimal time steps, the notion of position loses its meaning. We do not have one path anymore but all of them at once and the question of the position of the particle can only be answered statistically. The collection of all the paths can be thought of as a fluid whose density sets the chance probability of observing the particle in a given range of positions at a given time. It is that entire set that may then constitute the actual state of the quantum particle as becomes evident in the next section.

%%------------------------------------------------------------------------------------------------------------------------
%%------------------------------------------------------------------------------------------------------------------------
\section{Recovering wave-functions and Schr\"odinger's equation}\label{schrodinger}
  
The complex action $\mathcal S$\cite{nottale1993,nottale2011} can be re-expressed logarithmically in terms of a function $\psi$ with $\mathcal S=-i \mathcal S_0 \ln \left(\psi/\psi_0\right)$, in which $\psi_0$ and $\mathcal S_0$ are introduced for dimensional reasons. This can be used to express the complex velocity $\mathcal V$ by using the canonical momentum  $\mathcal P=m\mathcal V=\nabla \mathcal S$ or $\mathcal V=-i\frac{\mathcal S_0}{m}\nabla \ln \left(\psi/\psi_0\right)$. We see that $\psi_0$ cancels out of the expression of $\mathcal V$. For this reason and in order to lighten notations, we will start writing $\ln \psi$ in place of $\ln\left(\psi/\psi_0\right)$. This expression of $\mathcal V$ can be used together with the complex differential operator (Equation \ref{covardiff}) in the generalized fundamental relation of dynamics (Equation \ref{newton}), where we introduce $\eta=\frac{\mathcal S_0}{2m\mathcal D}$: 
$$ 2 i m\mathcal D\eta\left[{{\partial}\over{\partial t}}\left(\nabla\ln\psi\right) -i\mathcal D\left(2\eta(\nabla\ln\psi\nabla)(\nabla\ln\psi)    +\Delta(\nabla\ln\psi)\right)\right]=\nabla\Phi.$$

The identity\cite{nottale1993,nottale2011} demonstrated in Appendix \ref{identity} can be applied directly to obtain:
$$ 2im\mathcal D\eta\left[{{\partial}\over{\partial t}}\left(\nabla\ln\psi\right) -i{\mathcal D \over\eta}\nabla\left({{\Delta\psi^\eta}\over{\psi^\eta}}\right) \right]=\nabla\Phi.$$
Since all the terms are gradients, this can be integrated to
$$ 2im\mathcal D\eta{{\partial\ln\psi}\over{\partial t}} =-{2m\mathcal D^2 }\left({{\Delta\psi^\eta}\over{\psi^\eta}}\right)+\Phi+\Phi_0$$
where the integration constant $\Phi_0$ can always be absorbed in the choice of the origin of the energy scale so we do not carry it further. Developing the Laplacian and using $\mathcal P=-i\mathcal S_0\nabla\ln\psi=-2im\mathcal D\eta\nabla\ln\psi$, we obtain
$$ 2im\mathcal D\eta{{\partial\psi}\over{\partial t}} =\frac{\eta-1}{\eta}\frac{\mathcal P^2}{2m}\psi -2m\mathcal D^2\eta\Delta \psi +\Phi\psi.$$

If we now choose $\eta=1$, which corresponds to setting the value of the reference action $\mathcal S_0=2m\mathcal D$, we finally obtain Schr\"odinger's equation in which $\hbar$ is replaced with $2m\mathcal D$:
$$ 2im\mathcal D{{\partial\psi}\over{\partial t}} =-2m\mathcal D^2 \Delta\psi+\Phi\psi$$

This result is similar to that obtained by Edward Nelson as he "{\it examined the hypothesis that every particle of mass $m$ is subject to a Brownian motion with diffusion coefficient $\hbar/2m$ and no friction. The influence of an external field was expressed by means of Newton's law $F=ma$, $\cdots$}." 

However, the similarity is only superficial. Nelson concluded that "{\it the hypothesis leads in a natural way to Schr\"odinger's equation, but the physical interpretation is entirely classical\,}"\cite{Nelson1966}. Indeed, the diffusive process was postulated to be at play at some {\it sub-quantum} level, making it a hidden variable theory even if  "{\it the additional information which stochastic mechanics seems to provide, such as continuous trajectories, is useless, because it is not accessible to experimental verification}"\cite{Nelson1966}.  In the scale relativity approach  \cite{nottale1993,nottale2011} followed here, Schr\"odinger's equation does not result from any additional hypothesis. Instead, it results from the relaxation of the usually implicit hypothesis of differentiability for the space coordinates. Using the resolution-scale specific two-term path representation, we have seen that this relaxation corresponds to considering resolution-scales as additional relative attributes of reference frames, which is the central idea of resolution-scale relativity. The identification of the doubling of the velocity field in the two-term path representation led to the appearance of complex numbers \cite{nottale1993,nottale2011} and is in itself a definite departure from any {\it trajectory} based {\it classical interpretation}. With this, the generalized Newton relation and the equivalent Schr\"odinger equation take form under the specific restriction to  paths of fractal dimension $2$ corresponding to Wiener processes. Schr\"odinger's equation then appears as just one in a family of generally more intricate equations for stochastic processes with different statistics. 

Also, Nelson described quantum particles as having "{\it  continuous trajectories and the wave function is not a complete description of the state\,}"\cite{Nelson1966}. When discussing the simulation presented in Figure \ref{path}, we already commented on the fact that, in the limit of infinitesimal time steps, the position of the particle as a definite property becomes a meaningless concept, which has to be replaced with a probabilistic description. The consideration of {\it one specific path} being followed by a particle then is a misconception. Starting from the fundamental relation of dynamics generalized to non-differentiable paths, we arrive to Schr\"odinger's equation with the {\it wave function} $\psi({\bf x},t)$ identified to an exponential re-expression of the action. If the statistics of the stochastic component of the path is preserved all the way down to infinitesimal resolution time-scales, the state of the system can no longer be specified by coordinates values. With the disappearance of a specific path actually followed by the system, one must recognize the function $\psi$ as {\it completely} specifying the state of the system. 

Schr\"odinger's equation as a prescription for the time evolution of the state of the system is only one of the axioms founding quantum mechanics. In the following section, we discuss the interpretation of the other axioms in the scale relativistic approach. 
%%------------------------------------------------------------------------------------------------------------------------
%%------------------------------------------------------------------------------------------------------------------------
\section{The axioms of quantum mechanics}\label{axioms}
Standard quantum mechanics is built up from the enunciation of a number of mathematical postulates \cite{cohen1991}, which are generally not considered to derive from any more fundamental principles and are justified by the predictive power of their application. The first of these axioms specifies that the state of a system can be represented by a state-vector $|\psi\rangle$ belonging to a complex vectorial sate-space specific to the considered system. Another postulate, the Schr\"odinger postulate, prescribes the time evolution of the state of a system to be driven by Schr\"odinger's equation $i\hbar {{d}\over{dt}}|\psi\rangle=\hat H|\psi\rangle$, where $\hat H$ is the observable operator associated with the system total energy. The derivation of Schr\"odinger's equation in position representation (Section \ref{schrodinger}) identifies the wave-function $\psi(t,{\bf x})=\langle{\bf x}|\psi(t)\rangle$ (where $|\bf x\rangle$ is the state of definite position $\bf x$) to an exponential expression of the action $\mathcal S$, now a complex quantity in direct consequence of the consideration of non-differentiable paths. We already argued that the inclusion of non-differentiable paths amounts to a departure from a trajectory based description of the state of the system, leaving the wave function as the complete description of the state of the system. The time independent Schr\"odinger equation obtained by separation of variables is an eigenvalue equation. Its solutions constitute a vectorial space, which establishes the first postulate. In this section we discuss the other postulates from the scale relativistic point of view \cite{nottale2007}. 
%%------------------------------------------------------------------------------------------------------------------------
%%------------------------------------------------------------------------------------------------------------------------
\subsection{Observables as state-space operators}

The postulate on observables states that any physical quantity $\mathcal O$ that can be measured is associated with an hermitian operator $\hat O$ acting on the state space.  Such an operator is known as an observable. The measurement of a physical quantity can only yield one of the observable's eigenvalues $a_i$ as a result. 

In the course of the derivation of Schr\"odinger's equation, we have already identified the expression for the complex momentum $\mathcal P({\bf x})=\nabla \mathcal S=-i\mathcal S_0\nabla\ln\psi$ which can be rewritten as $\mathcal P({\bf x})\psi=-i\mathcal S_0\nabla\psi=\hat P\psi$. Similarly, for the energy we have $\mathcal E({\bf x})=\frac{\partial \mathcal S}{\partial t}=i\mathcal S_0\frac{\partial \ln\psi}{\partial t}$ or $\mathcal E({\bf x})\psi=i\mathcal S_0\frac{\partial\psi}{\partial t}=\hat E\psi$.  In both cases, the operator is found to be hermitian. When considering a state of definite momentum or energy, we may require $\mathcal P$ or $\mathcal E$ to be independent of $\bf x$. This then implies that the only possible values of a definite momentum or energy are the solutions of the usual eigenvalue equations. The correspondence between the possible outcomes of the measurements of physical quantities $\mathcal P$ and $\mathcal E$  and the respective linear operator $\hat P$ or $\hat E$ acting on the  state space is actually replaced by an equality.

More generally, in classical mechanics, any physical quantity characterizing the state of the system can be expressed as the result of some local operation on the classical action considered as a function of time and the system's coordinates. This may be generalized to the complex action $\mathcal S$ and, alternatively, we may consider the wave-function $\psi=\psi_0e^{i\mathcal S/\mathcal S_0}$ as a starting point. Then, any physical quantity $\mathcal O$  characterizing the state of the system can be expressed as the result of some local operation on the wave function $\psi$. Anticipating the wave function  $\psi(t,\bf x)$ as the complex probability amplitude of Born's postulate to be discussed in the next subsection (\ref{born}), implies $\mathcal O({\bf x})$ to be insensitive of the normalization and global phase of the wave function. This justifies the writing $\hat O\psi=\mathcal O({\bf x})\psi$ with  $\hat O$  a linear operator.  When considering a state of definite $\mathcal O$, we may require $\mathcal O({\bf x})$  to be independent of $\bf x$ and obtain an eigenvalue equation, which, as above, determines the only possible definite values of $\mathcal O$. The nature of the measurement process will be clarified in the discussion of von Neumann's postulate in subsection \ref{neuman}. We however already see how  regarding measurements outcomes as definite values of $\mathcal O$ implies they can only be eigenvalues of $\hat O$. In turn, the fact measurement outcomes are real quantities implies the observables $\hat O$ to be hermitian operators.
%%------------------------------------------------------------------------------------------------------------------------
%%------------------------------------------------------------------------------------------------------------------------
\subsection{Born's postulate}\label{born}
Born's postulate states that, for a system in a normalized state $|\psi\rangle$ (so that $\langle \psi|\psi\rangle=1$), the measurement of a quantity $\mathcal O$  yields one of the eigenvalues $o_i$ of the associated observable operator $\hat O$, with a chance probability given by the squared magnitude of the component of $|\psi\rangle$ in the sub-state-space corresponding to the observable eigenvalue $o_i$. In the case of position measurements, this postulate means that, in terms of the wave function $\psi(t,{\bf x})$, the probability density of finding the particle in $\bf x$ is given by $|\langle{\bf x}|\psi(t)\rangle|^2=|\psi(t,{\bf x})|^2$.

Writing the wave function as $\psi=\sqrt{\rho}\,e^{i\chi}$ in Schr\"odinger's equation established in Section \ref{schrodinger}, with both $\rho$ and $\chi$ real, and separating the real and imaginary parts, result in the Madelung \cite{madelung1927} equations (See \ref{madelung} for details): 

\begin{eqnarray}
{{\partial \rho}\over{\partial t}}&=&-\nabla\left(\rho {\bf V}\right)\label{madelungcont}\\
({{\partial}\over{\partial t}}+{\bf V}\nabla){\bf V}&=&-{{\nabla(\Phi+\mathcal Q)}\over{m}}\label{madelungeuler}
\end{eqnarray}

Equation \ref{madelungcont} is a continuity equation in which $\rho=\psi^*\psi$ plays the role of the fluid density with a velocity field ${\bf V}=\frac{\mathcal S_0}{m} \nabla \chi$ (See \ref{madelung}).  

Equation \ref{madelungeuler} is Euler's equation of fluid dynamics with the additional gradient of the quantum potential $\mathcal Q=-2m\mathcal D^2{{\Delta\sqrt{\rho}}\over{\sqrt{\rho}}}$ (See \ref{madelung}),  which appears to be entirely responsible for the quantum behavior. In the scale relativistic interpretation, the quantum potential is a manifestation of the fractal nature of the paths from which it derives. This is quite comparable to the situation in general relativity which reveals the gravitational potential as a manifestation of the curved nature of space-time \cite{Weinberg1972,Fock1964}.  

We already commented on the fact that abandoning the hypothesis of path differentiability results in the loss of path discernibility. If it were meaningful, following one path would amount to following them all. This restricts the consideration of position to a probabilistic description. Schr\"odinger's equation is now rewritten as a fluid dynamics equation. This naturally leads to identifying the fluid density $\rho=\psi\psi^*$ to the normalized density of indiscernible contributing paths. The normalized path density then sets the probability density of  position measurement outcomes. Indeed, the indiscernibility of the paths implies they are equally likely to be expressed in the measurement outcome. The time evolution of the system appears as a bundle of an infinity of indiscernable paths. The cross section of the bundle at a given time $t$ constitutes the state of the system and is described by the wave function $\psi(t,\bf x)$ solution of Schr\"odinger's equation. The normalized path density $\psi(t,{\bf x})\psi^*(t,{\bf x})$ of the bundle at time $t$, is the probability density for finding the particle in $\bf x$. 

This establishes Born's postulate in the position representation.  The wave function $\psi=\langle\bf x |\psi\rangle$ is just the position representation of an abstract state vector $|\psi\rangle$.  Observables, being Hermitian, the state can be represented in terms of the eigenstates of any complete set of commuting observables. That is to say, unitary transformations can be used to go from one representation to another. This generalizes Born's postulate in position representation to any representation.  The measurement of $\mathcal O$ gives $o_i$ with the probability $|\langle o_i|\psi\rangle|^2$ where $|o_i\rangle$ is a state of definite value $\mathcal O=o_i$, the eigenvector of $\hat O$ associated with the eigenvalue $o_i$.  
%%------------------------------------------------------------------------------------------------------------------------
%%------------------------------------------------------------------------------------------------------------------------
\subsection{von Neumann's postulate}\label{neuman}
The {\it  wave function collapse} or  von Neumann postulate specifies that, immediately after a measurement of $\mathcal O$ yielding $o_i$, the system is in the state given by the projection of the initial state onto the eigen-sub-state-space corresponding to the eigenvalue $o_i$ of the observable $\hat O$ corresponding to $\mathcal O$.

Given the above interpretation of Born's postulate with the wave functions describing the set of the non-differentiable and indiscernible paths, a measurement of $\mathcal O$ can naturally be envisioned as the selection of a bundle of indiscernible paths corresponding to the measurement outcome $o_i$, eigenvalue of  $\hat O$. The associated eigenstate vector $|o_i\rangle$ corresponds to the bundle of paths that is selected and for which $\mathcal O$ has the definite value $o_i$. It might be worth stressing again that the system should not be thought as following a specific path in the bundle. This would be a misconception  precisely because of the indiscernible character of the paths. The identification of the path bundle to the state of the system itself with the measurement amounting to a path bundle selection implies that the state of the system immediately after the measurement yielding $o_i$ precisely is $|o_i\rangle$. Following the initial measurement, after a time short enough for the time evolution of the selected bundle of paths to be negligible, a second measurement of $\mathcal O$ does not result in any further alteration of the paths bundle. The state vector remains $|o_i\rangle$ and the second measurement results in the same outcome $o_i$ with a unit probability.  This interpretation of von Neumann postulate follows the picture given for Born's postulate. The relative number of paths in the bundle $|o_i\rangle$ contributing to a state $|\psi\rangle$  is  $|\langle o_i|\psi\rangle|^2$, which sets the chance probability for the first measurement of $\mathcal O$ to yield $o_i$.
%%------------------------------------------------------------------------------------------------------------------------
%%------------------------------------------------------------------------------------------------------------------------
\subsection{Systems with more than one particle}
The previous subsection completes the scale relativistic interpretation of the postulates of standard quantum mechanics.  We went through the derivation of Schr\"odinger's equation for one particle in the usual three-dimensional physical space plus time. The exact same derivation can be carried out for an arbitrary number of dimensions. In particular, when considering a system composed of $N$ particles, we would obtain the same Schr\"odinger equation in a $3N$ dimensional physical space plus time, with, possibly, a different mass for each particle. The potential energy term may then depend on the relative coordinates of different particles to account for their mutual interactions.

In quantum mechanics, one talks about entanglement \cite{mermin1985} when two or more particles emerge from a mutual interaction in such a way one can only talk about the quantum state of the system as a whole and not about the quantum states of the constituents considered individually. The resulting correlation between the outcomes of the measurements of individual particles in the system is a major characteristic aspect of standard quantum mechanics \cite{schrodinger1935}. %Entanglement is in fact not specific to quantum mechanics. It is common in classical mechanics as there are quantities such as the energy, the momentum or the angular momentum that are globally conserved. Consequently, these quantities are not uniquely determined by any one specific constituent  of the system. This results in the fact that the measurement of the contribution of one constituent provides immediate knowledge about the contribution of the others, no matter how distant.  This, however, does not appear as a paradox as all the constituents have an actual position, to the contrary of the situation in quantum mechanics. 

In the scale relativity interpretation of quantum mechanics, an entangled state $|\psi_N\rangle$ of $N$ particles would correspond to a bundle of non-differentiable and indiscernible paths in the $3N$ dimensional physical space. The bundle may branch out in a number of sub-bundles corresponding to the various configurations in which the system might be found during a subsequent measurement of some of its constituent particles. Following the above interpretation of Born's and von Neumann's postulates, the statistics of the paths in the sub-bundles corresponds to the probabilities of the various possible outcomes of measurements of individual particles. We may stress again that the system of $N$ particles should not be thought as following one specific path. Instead, the state of the system is to be identified to the entire bundle of paths described by the state vector $|\psi_N\rangle$. The measurement of some particles then selects a part of the bundle and may immediately provide information about other particles, not involved in the measurement, because of the specific structure of the bundle which implements the appropriate correlation between the different parameters that may be measured in an experiment probing the entanglement.  

It appears that the early implementation of scale relativity in the development of point mechanics with time as an absolute external parameter lands onto a coherent foundation of standard quantum mechanics \cite{nottale1993,nottale2011}. This should at least be regarded as a validation of the scale relativity proposal. More has been done since with, in particular, a scale relativity approach to motion relativistic quantum mechanics \cite{celeriernottale2010} and also to gauge theories \cite{nottalecelerierlehner2006}.  Here, in the next section, we follow a different direction and consider the possibility that some macroscopic systems, which can be described as evolving along non-differentiable paths, could fall under a standard {\it quantum-like} description. 

%%------------------------------------------------------------------------------------------------------------------------
%%------------------------------------------------------------------------------------------------------------------------
\section{Chaos structured by a quantum-like mechanics}\label{chaos}

The above scale relativistic foundation of standard quantum mechanics does not result in any aspect different from other approaches that could be tested experimentally. It emerges from the consideration of non-differentiable stochastic paths described as Wiener processes all the way down to infinitesimal resolution-scales. It should however be noted that  the derivations of the generalized fundamental relation of dynamics and the equivalent Schr\"odinger equation (Section \ref{schrodinger}) do not depend in any way on the assumption that the fractal nature of  the paths is preserved uniformly all the way down to infinitesimal resolution scales. If there is a resolution scale below which the  paths loose their stochastic component, becoming differentiable and discernible again, Schr\"odinger's equation still holds and the other postulates are still applicable as long as the system is considered at sufficiently coarse resolution scales for the details of the evolution to be reducible to the statistical description of an effectively stochastic Wiener process. 

Chaotic systems are characterized by a high sensitivity to initial conditions \cite{gleik1987,thompson2016}.  This is generally described in terms of the rate at which two infinitesimally close trajectories move apart from each other, which defines the Lyapunov time-scales over which the chaotic nature of the dynamic system expresses itself. While the system may be evolving in a deterministic way, predictions of the evolution of the system over time intervals much exceeding the Lyapunov times are not reliable. This is often referred to as a predictability horizon. When the system is observed with resolution time-scales well in excess of the predictability horizon, the successive configurations appear random and uncorrelated. They sample an ensemble following a probability density map possibly evolving with time. The observation of the system with finer resolution-scales in attempts to better characterize an elusive trajectory keeps revealing new structures until one reaches the Lyapunov time-scale where the bundle of non-differentiable paths condensates into an actual differentiable trajectory and predictability is recovered.  So as long as the system is considered at resolution time-scales well exceeding the Lyapunov time, the developments that led us to a scale relativistic foundation of quantum mechanics should be applicable provided the resolution scale relativity principe is implemented in nature for such systems.

This has the intriguing consequence that the postulates of quantum mechanics may be applicable to macroscopic complex and/or chaotic systems outside the realm of standard quantum mechanics. The Planck constant in the Schr\"odinger equation would then be replaced by some different value $2m\mathcal D$ to be identified and which could be system specific.
Interestingly, the mass $m$ of the particle cancels out in the expression of the generalized de Broglie length $\lambda=\frac{2\mathcal D}{|\bf v|}$ and the velocity may then be expected to play a role similar to that of the momentum in standard quantum mechanics. Other than this, the main difference from standard quantum mechanics would lie in the fact that, at fine resolution-scale, a deterministic predictable behavior is recovered. So, in classical systems considered beyond their predictability horizon where classical mechanics fails and no alternative theory is currently available, a {\it quantum-like} mechanics may be applicable to provide some account for their often rich structuring. 

This justifies the search of {\it quantum-like} features in complex and chaotic  systems. Astrophysical systems with virtually only gravitation as an interaction force constitute a domain of predilection for such searches. In fact the possibility {\it quantum-like} structures could be found in Keplerian gravitational system was considered just a few years after the publication of Schr\"odinger's equation with an analysis of the orbits of the major objects of the Solar system as well as the orbits of their satellites\cite{caswell1929,malisoff1929,peniston1930}. 
These analyses were performed again in more details in the scale relativistic context \cite{nottale1997,hermann1998} and even included an account for the masses of the major objects of the Solar system following a {\it quantum-like} hydrogenoid orbital profile of the distribution of coalescing primordial planetesimals. Similar analyses were performed for Kuiper belt objects in the Solar System \cite{nottale2011}, extra-solar planetary systems\cite{nottale2000}, binary stars, pairs of galaxies and others \cite{nottale2011}.  All are suggestive that such a {\it quantum-like} mechanics is at play in the structuring of theses systems following multiples and submultiples of a seemingly universal velocity. While the compilation of these results is already striking, it would be highly desirable to achieve laboratory based experiment so the {\it quantum-like} dynamics with the emergence of characteristic velocities in the quantization can be tested in a controlled environment \cite{lebohec2017}.

%%------------------------------------------------------------------------------------------------------------------------
%%------------------------------------------------------------------------------------------------------------------------
\section{Conclusions and discussions}\label{conclusion}

The scale relativity proposal is to  include resolution-scales as additional relative parameters defining reference frames with respect to each others in an extension of the relativity principle. This  naturally  brings into focus fractal objects and, in Section \ref{fractal}, we reviewed the notion of fractal dimension which illustrated the requirement for the abandonment of the differentiability hypothesis set by the resolution scale relativity principle. In particular we considered the examples of Brownian and quantum mechanical paths, both found to be described by a fractal dimension $D_F=2$.

In Section \ref{differential}, we have exposed an approach to represent non-differentiable paths ${\bf q}(t)$ with two terms (Equation \ref{repres}). One describes the usual displacements revealed by inspection at some set resolution-scale and the other is the residual to be revealed at finer resolution-scales. We were led to recognize the velocity to be not only resolution-scale dependent but also double-valued, which can conveniently be expressed by making use of complex numbers (Equation \ref{complexv}). Correspondingly, we defined an also resolution-scale dependent and complex time-differential operator (Equation \ref{complex_diff_op}). Considering a regular field, $h({\bf x},t)$ along a non-differentiable path ${\bf q}(t)$, the complex time-differential operator acting on the field takes the form of a resolution-scale covariant differential (Equation \ref{covardiff}) defined under the restriction to paths in which the stochastic component is a Wiener process characterized by a diffusion constant $\mathcal D$. This correspond to a restriction to dynamical paths of fractal dimension $D_F=2$.

In Section \ref{lagrange}, applying the generalized stationary action principle, we established that the dynamics of non-differentiable paths is obtained by replacing the usual time derivative by the complex time-differential operator. In particular, we obtain a generalization of Newton's fundamental relation of dynamics (Equation \ref{newton}). 

Then, in Section \ref{harmonic}, we saw that, in the case of a zero average velocity $\langle \bf V\rangle=0$, the fundamental relation of dynamics takes the form of a Langevin equation, which we numerically integrated in the case of a simple harmonic oscillator. Guided by these numerical simulations, we understood that the adoption of non-differentiable dynamical paths corresponds to an abandonment of the notion of position or trajectory, which has to be replaced by an exclusively probabilistic consideration of position. We observed that the statistical distribution of the position of the path in the course of the numerical simulation reproduces the square of the magnitude of the solution of the time independent Schr\"odinger equation.  The quantum-like behavior appears to manifest itself as an emergence from the non-differentiability of the dynamical paths for the specific choice we made for the statistics of the stochastic term in the path representation.  The resulting scaling properties of the non-differentiable paths are expressed by the last term of the {\it scale-covariant time-differential operator} and are responsible for the quantum-like behavior. This is very similar to the situation in general relativity where the curved nature attributed to space time is expressed by the affine connection in the covariant derivative and is responsible for the gravitation phenomena. 

The connection with standard quantum mechanics is formalized in section \ref{schrodinger}, where, starting from the generalized equation of dynamics for a particle of mass $m$ (Equation \ref{newton}) and expressing the complex velocity $\mathcal V$ in terms of $\psi$, an exponential expression of the complex action $\mathcal S$, we obtained a Schr\"odinger equation with $\hbar$ replaced by $2m\mathcal D$.  We stress again that, because of their non-differentiability, the paths are not enumerable nor discernible in such a way it would be a misconception to envision a specific one to be followed by the system. Instead, the wave function $\psi$ must be recognized as completely specifying the state of the system as long as the fractal nature of the dynamical paths is preserved all the way to infinitesimal resolution-scales. As such, the wave function $\psi(t,\bf x)$ can be regarded as a description of the cross section of a bundle of an infinity of non-differentiable Wiener paths at time $t$.  
Standard quantum mechanics is observed at small resolution-scales where dynamical paths have a fractal dimensions $D_F=2$ while classical mechanics is recovered at larger resolution-scales where paths return to having fractal dimension $D_F=1$ and the bundle condensates into an actual trajectory. The transition between the two regimes occurs for resolution-scales comparable to the de Broglie wavelength. 

With this, in Section \ref{axioms}, we proceeded to a coherent scale relativistic interpretation of each of the postulates founding standard quantum mechanics. In particular, using Madelung's equations, the squared magnitude of the wave function was identified to the bundle's normalized paths density, which we assimilated to the modulus squared of the probability amplitude of Born's postulate. 

The establishment of Schr\"odinger's equation and the coherent interpretation of the postulates of quantum mechanics is a major success of scale relativity. This validation was continued with the application to relativistic\cite{celeriernottale2010} and gauge\cite{nottalecelerierlehner2006} quantum theories. Quantum mechanics is characterized by the fact it includes a dependance on resolution-scales as expressed most clearly by Heisenberg uncertainty relations. In hindsight, it is not surprising that scale relativity provides a natural and less axiomatic accommodation of quantum mechanics as the consideration for resolution-scale dependance is included from the very start by the extension of the relativity principle to resolution scaling laws. This success of the implementation of the resolution scale relativity principle may bring the question of the quantization of the gravitational interaction under a different light. The gravitational curvature of space-time at {\it large} scales may be seen as giving way to a dense structuring at small scales with non-differentiable and indiscernable geodesics in the quantum domain.  Scale relativity could provide an avenue for revealing quantum mechanics and the general relativistic description of gravitation, as being in continuation of each other.

In Section \ref{chaos}, we remarked that the preceding developments did not depend on the fractal nature of the considered paths to be uniformly preserved all the way down to infinitesimal resolution-scales. In particular, the postulates of quantum mechanics could remain applicable even if the paths loose their fractal character below some characteristic  resolution-scale.  Indeed, macroscopic systems may appear to evolve along  differentiable trajectories ($D_F=1$) when observed with a resolution finer than some mean-free-path length or Lyapunov scale. However, at much coarser resolutions, a description in terms of non-differentiable dynamical paths based on a Wiener process  ($D_F=2$) may become appropriate.  Provided the resolution scale relativity principle applies to this situation as well, this is the only requirement for the emergence of {\it quantum-like dynamics}. While the de Broglie wave length constitutes an upper bound for the resolution-scales at which quantum mechanics dominates, the mean-free-path length or Lyapunov time could constitute a lower bound for the resolution scale at which an emergent {\it quantum-like} mechanics would dominate. This opens the possibility for some complex and/or chaotic systems to be structured according to the laws of a {\it quantum-like} mechanics provided the systems are considered over resolution time-scales exceeding their predictability horizon. This is the domain in which classical mechanics looses its power, leaving probabilistic descriptions as the only valid approach while there is no currently accepted general tool or theory allowing for the prediction of probability densities. The observations of various astrophysical systems are already suggesting a {\it quantum-like} mechanics could be at play in their structures and dynamics \cite{caswell1929,malisoff1929,peniston1930,nottale1997,hermann1998,nottale2011,nottale2000} and therefore the resolution scale relativistic principle might indeed be implemented in nature. If it is the case, the application of the resolution scale relativity principle would provide a fruitful insight in complex/chaotic systems as their behavior is generally characterized by couplings across broad ranges of scales in a way which escapes the standard methods of physics. Additionally, it should be noted that the Schr\"odinger equation was obtained in the very common but specific case of paths with a Wiener process as their stochastic component. This corresponds to $D_F=2$ paths whose stochastic component is memoryless or Markovian. The consideration of different statistics would result in different dynamics, which maybe able to provide an account for structuring occurring in a broader range of complex natural systems.   
 
To summarize, it appears the resolution scale relativity principle provides a new approach to the foundation of quantum mechanics and may provide an effective method of theoretical research in the microphysical world. At the same time, it seems to provide an avenue to extend the reach of fundamental physics methods to integrate complex and chaotic systems, the approaches to which are otherwise restricted to effective and phenomenological descriptions. The program is ambitious and opens up on many possibilities of experimental, observational and theoretical developments in physics as well as in interdisciplinary fields as already illustrated in a number of recent publications inscribing the fundamental results outlined in this review in more general approaches \cite{calcagni2017,camelia2017,Chavanis2016} or applying them to problems in fields as varied as fundamental physics \cite{lebohec2017,Bhattacharya2017},  cosmology \cite{Chavanis2017}, atomic physics \cite{Duchateau2017}, material science \cite{turner2016a}, or biology \cite{turner2016b}.  

\section*{Acknowledgements}
The authors are grateful to Eugene Mishchenko, Patrick Fleury, and Janvida Rou for interesting discussions and their helpful comments and suggestions to clarify the text. 

\appendix
%%------------------------------------------------------------------------------------------------------------------------
\section{A useful identity}\label{identity}
%%------------------------------------------------------------------------------------------------------------------------
Let's look at $(\nabla\ln \psi)^2+\Delta\ln \psi=
\partial_i\ln \psi\partial_i\ln \psi+\partial_i\partial_i \ln \psi=
{{\partial_i \psi\partial_i \psi}\over{\psi^2}}+\partial_i{{\partial_i \psi}\over{\psi}}$ where $i=\{x,y,z\}$ with implicit summation over repeated indices. 
$(\nabla\ln \psi)^2+\Delta\ln \psi={{\partial_i f\partial_i \psi}\over{\psi^2}}+{{\psi\partial_i\partial_i \psi- \partial_i \psi\partial_i \psi}\over{\psi^2}}
={{\Delta \psi}\over{\psi}}$. We can then take the gradient: 
$$\nabla(\nabla \ln \psi)^2+\nabla\Delta\ln \psi=\nabla({{\Delta \psi}\over{\psi}}).$$

We now concentrate on the first term on the left hand side where we note $f=\ln\psi$: $\nabla(\nabla f)^2=\partial_i\partial_j f\partial_j f=2\partial_j f \partial_j \partial_i f$ so we get $$\nabla(\nabla f)^2=2(\nabla f \cdot \nabla)\nabla f.$$ 

So in total, using the fact that $\nabla\Delta=\Delta\nabla$, we can write: 
$$2(\nabla\ln\psi\cdot\nabla)\nabla\ln\psi+\Delta(\nabla\ln\psi)=\nabla\left({{\Delta\psi}\over{\psi}}\right).$$

Applying this to $\psi^\eta$ and dividing by $\eta$ we obtain:
$$2\eta(\nabla\ln\psi\cdot\nabla)\nabla\ln\psi+\Delta(\nabla\ln\psi)={1\over\eta}\nabla\left({{\Delta\psi^\eta}\over{\psi^\eta}}\right).$$
%%------------------------------------------------------------------------------------------------------------------------
\section{Madelung equations}\label{madelung}
%%------------------------------------------------------------------------------------------------------------------------

In position representation, the kinetic energy corresponds to the operator $\hat T = -\frac{\mathcal S_0}{2m}\Delta$ so that 
\begin{eqnarray}
\hat T \psi&=& -\frac{\mathcal S_0^2}{2m}\Delta\left(\sqrt{\rho}e^{i\chi}\right)\\
&=& -\frac{\mathcal S_0^2}{2m}e^{i\chi}\left(\Delta\sqrt{\rho}+2i\nabla\sqrt{\rho}\nabla\chi-\sqrt{\rho}\left(\nabla\chi\right)^2+i\sqrt{\rho}\Delta\chi\right)
\end{eqnarray}
Using $e^{i\chi}=\frac{\psi}{\sqrt\rho}$, simplifying by $\psi$ and rearranging a little, we obtain an expression for the kinetic energy: 
\begin{eqnarray}
\mathcal T &=& \frac{\mathcal S_0^2}{2m}\left(\nabla\chi\right)^2-\frac{\mathcal S_0^2}{2m}\frac{\Delta\sqrt{\rho}}{\sqrt\rho}-i\frac{\mathcal S_0^2}{2m\rho}\nabla\left(\rho\nabla\chi\right)\\
&=& \frac{1}{2}m{\bf V}^2+\mathcal Q-i\frac{\mathcal S_0}{2\rho}\nabla\left(\rho{\bf V}\right)\end{eqnarray}
In the second line we have introduced ${\bf V}=\frac{\mathcal S_0}{m}\nabla\chi$, the classical velocity field, with which the first term appears as the classical kinetic energy. It is worth noting that if we write $\chi=\frac{\mathcal S'}{2m\mathcal D}$, with $\mathcal S'$ the real part of the action, the expression of the velocity field corresponds the relation ${\bf V}=\frac{\nabla \mathcal S'}{m}$. As the gradient of the real part of the action equals the gradient of the classical action, this makes $\bf V$ clearly appear as a field of {\it usual velocity}. \\
We also introduced the potential $\mathcal Q=-\frac{\mathcal S_0^2}{2m}\frac{\Delta\sqrt\rho}{\sqrt\rho}$, the gradient of which was described by Erwin Madelung in his 1927 paper as "an internal force of the continuum" \cite{madelung1927}. The energy $\mathcal Q$ is now referred to as the {\it quantum potential}, a denomination due to Bohm in 1952\cite{bohm1952}. Despite this denomination and the role of $\mathcal Q$ in Madelung's equation, we see that the quantum potential  truly is of a kinetic nature \cite{holland2015}. 
 
We can proceed in the same way with the energy $\hat E\psi=i\mathcal S_0\frac{d}{dt}\left(\sqrt\rho e^{i\chi}\right)$, which leads to $\mathcal E =-\mathcal S_0\frac{d\chi}{dt}+i\frac{\mathcal S_0}{2\rho} \frac{d\rho}{dt}$. 

With the inclusion of the potential energy $\Phi$, the imaginary part of the equation $\mathcal E=\mathcal T+\Phi$  gives the continuity equation (Equation \ref{madelungcont}) while the gradient of the real part gives Euler's fluid dynamics equation (Equation \ref{madelungeuler}).

\end{document}